\documentclass[12pt, draftcls, onecolumn]{IEEEtran}
\usepackage{setspace}
\usepackage{amsmath, amsfonts, amssymb}
\usepackage{algorithmic}
\usepackage{algorithm}
\usepackage{array}
\usepackage[caption=false,font=normalsize,labelfont=scriptsize,textfont=scriptsize]{subfig}
\usepackage{textcomp}
\usepackage{stfloats}
\usepackage{url}
\usepackage{verbatim}
\usepackage{graphicx}
\usepackage{cite}
\usepackage{multirow}
\usepackage{dashrule}

\begin{document}

\title{Stacked Intelligent Metasurface Performs a 2D DFT in the Wave Domain for DOA Estimation}
\author{Jiancheng An, Chau Yuen, Marco Di Renzo, M\'erouane Debbah, H. Vincent Poor, and Lajos Hanzo
\thanks{J. An and C. Yuen are with the School of Electrical and Electronics Engineering, Nanyang Technological University, Singapore 639798 (e-mail: jiancheng\_an@163.com; chau.yuen@ntu.edu.sg). M. Di Renzo is with CNRS, CentraleSup\'elec, Laboratoire des Signaux et Syst\`emes, Universit\'e Paris-Saclay, 91192 Gif-sur-Yvette, France (e-mail: marco.di-renzo@universite-paris-saclay.fr). M. Debbah is with the Center for 6G Technology, Khalifa University of Science and Technology, P O Box 127788, Abu Dhabi, United Arab Emirates (e-mail: merouane.debbah@ku.ac.ae). H. Vincent Poor is with the Department of Electrical and Computer Engineering, Princeton University, Princeton, NJ 08544 USA (e-mail: poor@princeton.edu). L. Hanzo is with the School of Electronics and Computer Science, University of Southampton, SO17 1BJ Southampton, U.K. (e-mail: lh@ecs.soton.ac.uk).}\vspace{-0.9cm}}
\markboth{DRAFT}{DRAFT}
\maketitle
\begin{abstract}
Staked intelligent metasurface (SIM) based techniques are developed to perform two-dimensional (2D) direction-of-arrival (DOA) estimation. In contrast to the conventional designs, an advanced SIM in front of the receiving array automatically performs the 2D discrete Fourier transform (DFT) as the incident waves propagate through it. To arrange for the SIM to carry out this task, we design a gradient descent algorithm for iteratively updating the phase shift of each meta-atom in the SIM to minimize the fitting error between the SIM's response and the 2D DFT matrix. To further improve the DOA estimation accuracy, we configure the phase shifts in the input layer of SIM to generate a set of 2D DFT matrices having orthogonal spatial frequency bins. Extensive numerical simulations verify the capability of a well-trained SIM to perform 2D DFT. Specifically, it is demonstrated that the SIM having an optical computational speed achieves an MSE of $10^{-4}$ in 2D DOA estimation.
\end{abstract}


\section{Introduction}
\IEEEPARstart{D}{irection-of-arrival} (DOA) estimation has long been a crucial task in applications areas, such as astronomy, navigation, and radar sensing systems \cite{TAP_1986_Schmidt_Multiple, TASSP_1989_Roy_ESPRIT, TASSP_1989_Stoica_MUSIC, SPL_2020_Zheng_Direction, TWC_2023_An_Fundamental}. Traditionally, the DOAs have been estimated using beamforming relying on the fast Fourier transform (FFT) technique \cite{BOOK_1985_Haykin_Array, SPM_1996_Krim_Two}. However, beamforming has a limited angular resolution, which is fundamentally restricted by the Rayleigh criterion \cite{Proc_1997_Godara_Application}. To address this problem, several super-resolution DOA estimation approaches have been developed such as MUSIC and ESPRIT \cite{TAP_1986_Schmidt_Multiple, TASSP_1989_Roy_ESPRIT, TASSP_1989_Stoica_MUSIC}. While these methods provide significant performance advantages, they require potentially excessive computational and storage resources to carry out the eigenvalue decomposition of the spatial covariance matrix. Additionally, their performance erodes when only a small number of snapshots are available.

Moreover, practical systems generally face various imperfections such as station location errors, which are impervious to accurate modeling and calibration using conventional methods \cite{TSP_2004_See_Direction, TSP_2013_Liu_A}. Fortunately, advanced machine learning (ML) techniques are capable of accurately estimating the DOAs. In contrast to conventional model-based methods, data-driven ML-based approaches are more robust against practical array imperfections. Nevertheless, traditional DNNs rely on commercial processors or dedicated chips to perform computations, whose speed is limited by digital hardware. Motivated by advances in metasurface technologies and the radical wave-based computing paradigm \cite{TVT_2023_Xu_Channel, TGCN_2022_An_Joint, JSAC_2020_Di_Smart, TCOM_2022_An_Low, Light_2014_Cui_Coding}, Liu \emph{et al.} \cite{NE_2022_Liu_A} fabricated a reconfigurable diffractive deep neural network (D$^2$NN) using a stacked intelligent metasurface (SIM) for large-scale parallel calculations and analog signal processing to occur at the speed of light \cite{Science_2014_Silva_Performing}. Specifically, a SIM employs an array of programmable metasurface layers \cite{WC_2022_An_Codebook}, each containing many programmable meta-atoms that can manipulate the EM behavior, as the waves pass through it. Adapting the bias voltage via a customized field-programmable gate array (FPGA) module allows each meta-atom to act as a reprogrammable artificial neuron having tunable weights.

Inspired by this advanced architecture, the authors of \cite{JSAC_2023_An_Stacked} harnessed SIM to implement holographic multiple-input multiple-output (MIMO) communications \cite{CL_2023_An_A}. In contrast to conventional MIMO designs, a pair of SIMs deployed at the transmitter and receiver can automatically accomplish MIMO transmit precoding and receive combining as the EM waves propagate through them. This allows each spatial stream to be directly radiated and recovered from its corresponding transmit and receive ports, while significantly reducing the number of radio frequency (RF) chains needed. Furthermore, the authors of \cite{ICC_2023_An_Stacked, arXiv_2023_An_Stacked} harnessed a SIM at the BS to facilitate downlink multiuser beamforming in the EM wave domain. This eliminates the need for conventional digital beamforming and for high-resolution digital-to-analog converters (DACs) at the BS.

Nevertheless, DOA estimation using an advanced SIM remains unexplored. Against this background, we design a new SIM-based physical DOA estimator in this paper. The philosophy behind the proposed estimator is that by optimizing the SIM, the incident EM waves can be transformed into the spatial frequency domain as they propagate through it. As a result, the signal's direction can be readily estimated by detecting the energy levels at different receiver probes. This SIM-based system concept substantially simplifies the receiver hardware, since no analog-to-digital converters (ADC) are required. Most remarkably, the calculation in the SIM occurs naturally as the waves propagate through it without incurring a processing delay.

\section{SIM-Based Array System Model}\label{sec2}
In this section, we present the system model for the SIM-based array used for DOA estimation.
\subsection{Incident Signal Model}
\begin{figure*}[!t]
\centering
\includegraphics[width=17cm]{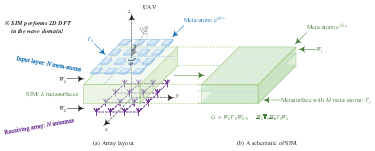}
\caption{Illustration of a SIM-based radar system, where the SIM performs 2D DFT in the wave domain. The receiving array directly observes the spatial spectrum of the incident signal.}\vspace{-0.6cm}
\label{fig_1}
\end{figure*}
As depicted in Fig. \ref{fig_1}, we utilize a uniform planar array (UPA) placed on the ground (i.e., the $x$-$y$ plane) for estimating the DOA parameters. In contrast to conventional array systems, a SIM consisting of $\left ( L+1 \right )$ metasurface layers is integrated with the UPA to transform the incident signal into its spatial frequency domain. We assume that the SIM is positioned horizontally, with all metasurface layers parallel to the $x$-$y$ plane. To avoid ambiguity, the metasurfaces are labeled by $0 \sim L$ from the top to the bottom, as shown in Fig. \ref{fig_1}. Let $\varphi \in \left [ 0,2\pi \right )$ and $\vartheta \in \left [ 0,\pi/2 \right ]$ represent the physical azimuth angle and elevation angle of the DOA of the radiation source relative to the input layer of the SIM, which has $N = N_{\textrm{x}} N_{\textrm{y}}$ meta-atoms, with $N_{\textrm{x}}$ and $N_{\textrm{y}}$ representing the number of meta-atoms in the $x$- and $y$-directions, respectively. Additionally, the corresponding element spacings are $d_{\textrm{x}}$ and $d_{\textrm{y}}$. Therefore, the electrical angles $\psi _{\textrm{x}}$ and $\psi _{\textrm{y}}$ in the $x$- and $y$-directions are given by \cite{TSP_2012_Heidenreich_Joint}
\begin{align}
 \psi _{\textrm{x}}&=\kappa d_{\textrm{x}}\sin\left ( \vartheta \right )\cos\left ( \varphi \right ), \label{eq1}\\
\psi _{\textrm{y}}&=\kappa d_{\textrm{y}}\sin\left ( \vartheta \right )\sin\left ( \varphi \right ), \label{eq2}
\end{align}
respectively, where $\kappa =2\pi /\lambda $ represents the wavenumber, with $\lambda$ being the wavelength.

Hence, the steering vector $\boldsymbol{a}\left ( \psi _{\textrm{x}}, \psi _{\textrm{y}} \right )\in \mathbb{C}^{N \times 1}$ \emph{w.r.t.} the input layer of the SIM is written as
\begin{align}
\boldsymbol{a}\left ( \psi _{\textrm{x}}, \psi _{\textrm{y}} \right )=\boldsymbol{a}_{\textrm{y}}\left ( \psi _{\textrm{y}} \right ) \otimes \boldsymbol{a}_{\textrm{x}}\left ( \psi _{\textrm{x}} \right ), \label{eq3}
\end{align}
and the elements of the vectors $\boldsymbol{a}_{\textrm{x}}\left ( \psi _{\textrm{x}} \right )\in \mathbb{C}^{N_{\textrm{x}}\times 1}$ and $\boldsymbol{a}_{\textrm{y}}\left ( \psi _{\textrm{y}} \right )\in \mathbb{C}^{N_{\textrm{y}}\times 1}$ are defined as follows:
\begin{align}
\left [ \boldsymbol{a}_{\textrm{x}}\left ( \psi _{\textrm{x}} \right ) \right ]_{n_{\textrm{x}}}&\triangleq e^{j\psi _{\textrm{x}}\left ( n_{\textrm{x}}-1 \right )},\quad n_{\textrm{x}}=1,2,\cdots ,N_{\textrm{x}}, \label{eq4}\\
 \left [ \boldsymbol{a}_{\textrm{y}}\left ( \psi _{\textrm{y}} \right ) \right ]_{n_{\textrm{y}}}&\triangleq e^{j\psi _{\textrm{y}}\left ( n_{\textrm{y}}-1 \right )},\quad n_{\textrm{y}}=1,2,\cdots ,N_{\textrm{y}}. \label{eq5}
\end{align}

Let $s\in \mathbb{C}$ represent the signal transmitted from the radiation source, which is modeled as a circularly symmetric complex Gaussian (CSCG) random variable with zero mean and unit variance. Hence, the signal $\boldsymbol{x}\in \mathbb{C}^{N\times 1}$ incident upon the input layer of the SIM can be expressed as
\begin{align}
 \boldsymbol{x}=\boldsymbol{a}\left ( \psi _{\textrm{x}}, \psi _{\textrm{y}} \right )s. \label{eq6}
\end{align}
\subsection{SIM Model}
Fig. \ref{fig_1}(b) shows the schematic diagram of a SIM device. For the sake of brevity, we assume that each of the $L$ intermediate metasurface layers is modeled as a UPA obeying isomorphic arrangements. Additionally, we assume that the $\left ( L+1 \right )$ metasurfaces are evenly spaced. Let $T_{\textrm{SIM}}$ represent the thickness of the SIM. As such, the vertical spacing between adjacent layers is obtained by $s_{\textrm{layer}}=T_{\textrm{SIM}}/L$. In practice, the SIM is enclosed in a supporting structure surrounded by wave-absorbing material to reduce the interference from undesired diffraction and scattering \cite{NE_2022_Liu_A}. Furthermore, each metasurface layer consists of $M = M_{\textrm{x}} M_{\textrm{y}}$ meta-atoms, where $M_{\textrm{x}}$ and $M_{\textrm{y}}$ are the number of meta-atoms in the $x$- and $y$-directions, respectively. Moreover, the corresponding spacings between the adjacent meta-atoms on the intermediate layers are set to $s_{\textrm{x}}$ and $s_{\textrm{y}}$.

Moreover, each meta-atom is capable of adjusting the phase shift of the EM waves passing through it by controlling the bias voltage of the associated circuit \cite{JSAC_2023_An_Stacked, arXiv_2023_An_Stacked, NE_2022_Liu_A}. Let $\boldsymbol{\upsilon}_{l}=\left [\upsilon_{l,1},\upsilon_{l,2},\cdots ,\upsilon_{l,M} \right ]^{T}\in \mathbb{C}^{M\times 1},\ l = 1,2,\cdots ,L$ represent the complex-valued transmission coefficient vector for the $l$-th layer, where $\upsilon_{l,m}=e^{j\xi _{l,m}},\ m=1,2, \cdots, M,\ l=1,2, \cdots, L$ with $ \xi _{l,m}\in \left [ 0,2\pi \right )$ representing the phase shift of the $m$-th meta-atom on the $l$-th layer \cite{JSAC_2023_An_Stacked}. Furthermore, let $\boldsymbol{\Upsilon}_{l} =\textrm{diag}\left ( \boldsymbol{\upsilon }_{l} \right )\in \mathbb{C}^{M\times M}$ represent the corresponding transmission coefficient matrix for the $l$-th layer. In particular, let $\boldsymbol{\upsilon}_{0}=\left [\upsilon_{0,1},\upsilon_{0,2},\cdots ,\upsilon_{0,N} \right ]^{T}\in \mathbb{C}^{N\times 1}$ and $\boldsymbol{\Upsilon}_{0} =\textrm{diag}\left ( \boldsymbol{\upsilon }_{0} \right )\in \mathbb{C}^{N\times N}$ denote the complex-valued transmission coefficient vector and the corresponding matrix for the input layer, respectively, where we have $\upsilon_{0,n}=e^{j\xi _{0,n}},\ n=1,2, \cdots, N$ and $\xi _{0,n}$ denotes the phase shift of the $n$-th meta-atom on the input layer.

Furthermore, let $\boldsymbol{W}_{l}\in \mathbb{C}^{M\times M},\ l = 1,2,\cdots ,L-1$ characterize the EM wave propagation between adjacent layers in SIM. Based on the Rayleigh-Sommerfeld diffraction equation \cite{JSAC_2023_An_Stacked}, the attenuation coefficient $\left [ \boldsymbol{W}_{l} \right ]_{m,\breve{m}}$ from the $\breve{m}$-th meta-atom on layer $l$ to the $m$-th meta-atom on layer $\left ( l+1 \right )$  is formulated as follows:
\begin{align}
\left [ \boldsymbol{W}_{l} \right ]_{m,\breve{m}}=\frac{A_{\textrm{meta-atom}}s_{\textrm{layer}}}{2\pi d_{m,\breve{m}}^{3}}\left ( 1-j\kappa d_{m,\breve{m}}\right )e^{j \kappa d_{m,\breve{m}}},\label{eq7}
\end{align}
where $A_{\textrm{meta-atom}}$ denotes the area of each meta-atom, and $d_{m,\breve{m}}$ represents the corresponding propagation distance, which is calculated as follows:
\begin{align}
 d_{m,\breve{m}}=\sqrt{\left ( m_{\textrm{x}}-\breve{m}_{\textrm{x}} \right )^{2}s_{\textrm{x}}^{2}+\left ( m_{\textrm{y}}-\breve{m}_{\textrm{y}} \right )^{2}s_{\textrm{y}}^{2}+s_{\textrm{layer}}^{2}}, \label{eq8}
\end{align}
with $m_{\textrm{y}}$ and $m_{\textrm{x}}$ being defined by $m_{\textrm{y}} \triangleq \left \lceil m/M_{\textrm{x}} \right \rceil$ and $m_{\textrm{x}} \triangleq m-\left ( m_{\textrm{y}}-1 \right )M_{\textrm{x}}$. Similarly, $\breve{m}_{\textrm{y}}$ and $\breve{m}_{\textrm{x}}$ are obtained by replacing $m$ with $\breve{m}$.

Furthermore, let $\boldsymbol{W}_{0}\in \mathbb{C}^{M\times N}$ represent the attenuation coefficient matrix between the input layer and the first layer. The $\left ( m,n \right )$-th entry of $\boldsymbol{W}_{0}$ is obtained by replacing $d_{m,\breve{m}}$ in \eqref{eq7} with the corresponding propagation distance $\tilde{d}_{m,n}$. The value of $\tilde{d}_{m,n}$ can be readily calculated according to the array layout seen in Fig. \ref{fig_1}(a). Similarly, let $\boldsymbol{W}_{L}\in \mathbb{C}^{N\times M}$ represent the attenuation coefficient matrix between the $L$-th metasurface layer and the output layer, i.e., the receive antenna array. The receiver is a UPA arranged in the same pattern as the input layer of SIM, and it is placed at $s_{\textrm{layer}}$ away from the $L$-th layer. Because of the identical layer spacing and the isomorphic arrangement of the input and output layers, it is readily seen that $\boldsymbol{W}_{L} = \boldsymbol{W}_{0}^{T}$.

As a result, the overall forward propagation through the SIM $\boldsymbol{G}\in \mathbb{C}^{N\times N}$ is described as:
\begin{align}
\boldsymbol{G}=\boldsymbol{W}_{L}\boldsymbol{\Upsilon} _{L}\boldsymbol{W}_{L-1}\cdots \boldsymbol{W}_{2}\boldsymbol{\Upsilon} _{2}\boldsymbol{W}_{1}\boldsymbol{\Upsilon} _{1}\boldsymbol{W}_{0}. \label{eq15}
\end{align}
\subsection{Received Signal Model}
As mentioned earlier, the receiver has a UPA consisting of $N = N_{\textrm{x}} N_{\textrm{y}}$ receiver antennas. For a single source transmitting a waveform $s$, the complex signal vector $\boldsymbol{r}\in \mathbb{C}^{N\times 1}$ received at the array can be expressed as:
\begin{align}
\boldsymbol{r}=\sqrt{\varrho }\boldsymbol{G}\boldsymbol{\Upsilon} _{0}\boldsymbol{x}+\boldsymbol{u}=\sqrt{\varrho }\boldsymbol{G}\boldsymbol{\Upsilon} _{0}\boldsymbol{a}\left ( \psi _{\textrm{x}},\psi _{\textrm{y}} \right )s+\boldsymbol{u}, \label{eq16}
\end{align}
where $\varrho$ represents the signal-to-noise ratio (SNR), and $\boldsymbol{u}\in \mathbb{C}^{N\times 1}$ is the measurement noise vector at the receiving array, which is modeled as a CSCG random vector satisfying $\boldsymbol{u}\sim \mathcal{CN}\left ( \boldsymbol{0}, \boldsymbol{I}_{N} \right )$. It is also assumed that $s$ and $\boldsymbol{u}$ are uncorrelated.

Note that in contrast to the conventional array, the received signal in \eqref{eq16} is processed by a controllable analog transformation carried out by the SIM. By appropriately configuring the phase shifts in the SIM, the receiving antenna array has the potential to directly observe the spatial-domain spectrum of the incident signal.

\section{SIM Optimization for Realizing 2D DFT}\label{sec3}
Achieving the expected DOA estimation capability requires the SIM to output the spatial-domain spectrum of the incoming signal. In this section, we formulate an optimization problem for using the SIM to implement the 2D DFT in the wave-domain and devise a gradient descent algorithm for finding a high-quality suboptimal solution for the phase shifts.

\subsection{Optimization Problem}
Specifically, for the 2D DFT matrix $\boldsymbol{F}\in \mathbb{C}^{N\times N}$ having $N = N_{\textrm{x}}N_{\textrm{y}}$ grid points, its $\left ( n,\breve{n} \right )$-th entry is defined as follows:
\begin{align}
 f_{n,\breve{n}} = \left [ \boldsymbol{F} \right ]_{n,\breve{n}}\triangleq e^{-j2\pi \frac{\left ( n_{\textrm{x}}-1 \right )\left ( \breve{n}_{\textrm{x}}-1 \right )}{N_{\textrm{x}}}}e^{-j2\pi \frac{\left ( n_{\textrm{y}}-1 \right )\left ( \breve{n}_{\textrm{y}}-1 \right )}{N_{\textrm{y}}}}, \label{eq17}
\end{align}
where $n_{\textrm{y}}$ and $n_{\textrm{x}}$ are defined as $n_{\textrm{y}} \triangleq \left \lceil n/N_{\textrm{x}} \right \rceil$ and $n_{\textrm{x}} \triangleq n-\left ( n_{\textrm{y}}-1 \right )N_{\textrm{x}}$, respectively, while $\breve{n}_{\textrm{y}}$ and $\breve{n}_{\textrm{x}}$ are obtained by replacing $n$ with $\breve{n}$.

To evaluate the similarity between the SIM's response and the 2D DFT matrix, the loss function $\mathcal{L}$ is defined as the Frobenius norm of the fitting error between the target entry and the EM response of the SIM, yielding
\begin{align}
 \mathcal{L} = \left \| \beta \boldsymbol{G}-\boldsymbol{F} \right \|_{F}^{2}, \label{eq18}
\end{align}
where $\beta \in \mathbb{C}$ represents the scaling factor required for helping the SIM's response at the normalized value required.

Aiming at minimizing the loss function in \eqref{eq18}, the optimization problem constructed for utilizing the SIM to fit the 2D DFT matrix is formulated as:
\begin{subequations}\label{eq19}
\begin{alignat}{2}
&\!\min_{\left \{ \xi _{l,m} \right \}} &\quad & \mathcal{L} = \left \| \beta \boldsymbol{G}-\boldsymbol{F} \right \|_{F}^{2} \label{eq19a}\\
&\textrm{s.t.} & & \boldsymbol{G}=\boldsymbol{W}_{L}\boldsymbol{\Upsilon} _{L}\boldsymbol{W}_{L-1}\cdots \boldsymbol{W}_{2}\boldsymbol{\Upsilon} _{2}\boldsymbol{W}_{1}\boldsymbol{\Upsilon} _{1}\boldsymbol{W}_{0}, \label{eq19b}\\
& & & \boldsymbol{\Upsilon}_{l}=\textrm{diag}\left ( \left [e^{j\xi _{l,1}},e^{j\xi _{l,2}},\cdots ,e^{j\xi _{l,M}} \right ]^{T}\right ), \label{eq19c}\\
& & & \xi _{l,m} \in \left [ 0,2\pi \right ),\, m=1, \cdots, M,\, l=1, \cdots, L, \label{eq19d}\\
& & & \beta \in \mathbb{C}. \label{eq19e}
\end{alignat}
\end{subequations}

Note that due to the non-convex constant modulus constraint and the fact that the phase shifts associated with different metasurface layers are highly coupled, the optimization problem in \eqref{eq19} is non-trivial to solve. In the subsequent subsection, we customize a gradient descent method for efficiently finding a near-optimal solution of \eqref{eq19}.
\subsection{Proposed Gradient Descent Algorithm}\label{sec3_2}
The gradient descent algorithm iteratively adjusts the phase shifts in the SIM for minimizing the loss function in \eqref{eq19a}. Specifically, the gradient descent involves two main procedures: \emph{i)} calculating the derivative and \emph{ii)} updating the parameters.
\subsubsection{Derivative Calculation}
For a tentative SIM model, the gradient of the loss function $\mathcal{L}$ \emph{w.r.t.} the phase shift vector $\boldsymbol{\xi} _{l}$ of the $l$-th layer in a SIM is calculated by:
\begin{align}
\nabla_{\boldsymbol{\xi} _{l}} \mathcal{L} &=\sum_{n=1}^{N}\nabla_{\boldsymbol{\xi} _{l}} \left \|\beta \boldsymbol{g}_{n}-\boldsymbol{f}_{n} \right \|^{2},\ l=1,2,\cdots ,L. \label{eq43}
\end{align}
where $\boldsymbol{g}_{n}\in \mathbb{C}^{N\times 1},\ n=1, 2, \cdots, N$ and $\boldsymbol{f}_{n}\in \mathbb{C}^{N\times 1},\ n=1, 2, \cdots, N$ represent the $n$-th column of $\boldsymbol{G}$ and $\boldsymbol{F}$, respectively.

Furthermore, the $m$-th entry of the gradient in \eqref{eq43} is obtained by taking the partial derivative of $\left \|\beta \boldsymbol{g}_{n}-\boldsymbol{f}_{n} \right \|^{2}$ \emph{w.r.t.} $\xi _{l,m}$. Upon applying the chain rule to the derivatives we have:
\begin{align}\label{eq44}
\frac{\partial \left \|\beta \boldsymbol{g}_{n}-\boldsymbol{f}_{n} \right \|^{2} }{\partial \xi _{l,m}}&=2\Re\left \{ \beta^{\ast } \frac{\partial \boldsymbol{g}_{n}^{H}}{\partial \xi _{l,m}}\left ( \beta \boldsymbol{g}_{n}-\boldsymbol{f}_{n} \right ) \right \} \notag\\
&\overset{\left ( i \right )}{=}2\Re\left \{ \beta^{\ast } \frac{\partial \left (\boldsymbol{P}_{l,n}\boldsymbol{\upsilon} _{l}\right )^{H}}{\partial \xi _{l,m}}\left ( \beta \boldsymbol{g}_{n}-\boldsymbol{f}_{n} \right ) \right \} \notag\\
&=2\Re\left \{ \beta^{\ast } \frac{1}{j}\upsilon _{l,m}^{\ast }\boldsymbol{e}_{m}^{H}\boldsymbol{P}_{l,n}^{H}\left ( \beta \boldsymbol{g}_{n}-\boldsymbol{f}_{n} \right ) \right \} \notag\\
&=2\Im\left \{ \beta^{\ast } \upsilon _{l,m}^{\ast }\boldsymbol{e}_{m}^{H}\boldsymbol{P}_{l,n}^{H}\left ( \beta \boldsymbol{g}_{n}-\boldsymbol{f}_{n} \right ) \right \},
\end{align}
for $m=1,2,\cdots ,M,\ l=1,2,\cdots ,L$, where $\boldsymbol{e}_{m}^{H}$ represents the $m$-th row of the identity matrix $\boldsymbol{I}_{M}$ and $\left ( i \right )$ holds due to the fact that $\boldsymbol{g}_{n}=\boldsymbol{P}_{l,n}\boldsymbol{\upsilon} _{l}$, and $\boldsymbol{P}_{l,n}\in \mathbb{C}^{N\times M},\ n=1, 2, \cdots, N,\ l=1, 2, \cdots, L$ denotes the equivalent coefficient matrix associated with the $l$-th metasurface layer activating the $n$-th meta-atom on the input layer, which is defined as
\begin{align}
 \boldsymbol{P}_{l,n}= \boldsymbol{W}_{L}\boldsymbol{\Upsilon} _{L}\boldsymbol{W}_{L-1}\cdots \boldsymbol{W}_{l+1}\boldsymbol{\Upsilon} _{l+1}\boldsymbol{W}_{l}\textrm{diag}\left ( \boldsymbol{q}_{l,n} \right ),\label{eq21}
\end{align}
with $\boldsymbol{q}_{l,n}\in \mathbb{C}^{M\times 1}$ representing the complex signal component illuminating the $l$-th layer of the SIM from the $n$-th meta-atom in the input layer, defined as
 \begin{align}
\boldsymbol{q}_{l,n}&=\boldsymbol{W}_{l-1}\boldsymbol{\Upsilon} _{l-1}\boldsymbol{W}_{l-2}\cdots \boldsymbol{W}_{2}\boldsymbol{\Upsilon} _{2}\boldsymbol{W}_{1}\boldsymbol{\Upsilon} _{1}\boldsymbol{w}_{0,n},
\end{align}
for $n=1, 2, \cdots, N,\ l = 1,2,\cdots ,L$, where $\boldsymbol{w}_{0,n}\in \mathbb{C}^{M\times 1}$ represents the $n$-th column of $\boldsymbol{W}_{0}$.

By gathering the $M$ partial derivatives in \eqref{eq44} into a vector, the gradient in \eqref{eq43} can be formulated as:
\begin{align}
\nabla_{\boldsymbol{\xi} _{l}} \mathcal{L} &=2\sum_{n=1}^{N}\Im\left \{ \beta ^{\ast } \boldsymbol{\Upsilon} _{l}^{H}\boldsymbol{P}_{l,n}^{H}\left ( \beta \boldsymbol{g}_{n}-\boldsymbol{f}_{n} \right ) \right \}, \label{eq20}
\end{align}
for $l=1, 2, \cdots, L$. 

\subsubsection{Parameter Update}
Once all the gradients \emph{w.r.t.} the SIM's phase shift vectors have been calculated, we simultaneously evolve the phase shift values $\boldsymbol{\xi} _{l}$ in the direction that decreases the loss function value. At each iteration, the update formula is as follows:
\begin{align}
\boldsymbol{\xi }_{l}\leftarrow \boldsymbol{\xi }_{l}-\eta \nabla_{\boldsymbol{\xi }_{l}}\mathcal{L}, \label{eq23}
\end{align}
where $\eta >0$ represents the learning rate. To ensure a stable convergence, the learning rate $\eta$ also decreases over time according to a step-based schedule:
\begin{align}
    \eta\leftarrow \eta \zeta \pi /\underset{l=1,2,\cdots ,L}{\max}\left \{ \max \nabla_{\boldsymbol{\xi }_{l}}\mathcal{L} \right \}, \label{eq24}
\end{align}
with $\zeta $ representing the step-size decay parameter.

In addition, the auxiliary scaling factor $\beta$ also has to be updated during each iteration to maintain the normalized level. Specifically, given a tentative SIM response matrix $\boldsymbol{G}$, the optimal value of $\beta$ can be readily obtained by utilizing the least squares (LS) method, yielding
\begin{align}
 \beta = \left ( \boldsymbol{g} ^{H}\boldsymbol{g} \right )^{-1} \boldsymbol{g}^{H}\boldsymbol{f},
\end{align}
where we have $\boldsymbol{g} = \textrm{vec}\left ( \boldsymbol{G} \right )$ and $\boldsymbol{f} = \textrm{vec}\left ( \boldsymbol{F} \right )$.

The phase shift values are updated repeatedly until either the loss function $\mathcal{L}$ converges or the SIM's phase shift vectors have been updated for the affordable number of iterations. By employing the SIM to implement the 2D DFT in the wave domain, the system can directly generate the spatial-domain spectrum at the receiving array and provide a coarse on-grid estimate for the DOA parameters of the incident signal. However, the coarse estimate has limited estimation precision for a small value of $N$. Fortunately, the input layer provides an extra design degree of freedom (DoF), which can be exploited for generating a set of 2D DFT matrices associated with different frequency bins, thus substantially improving the DOA estimation performance of a moderate-size SIM, as it will be further detailed in the next section.

\section{SIM-Based DOA Estimation}\label{sec4}
In this section, we first introduce the proposed SIM-based DOA estimation protocol by appropriately configuring the phase shift values of the input layer, i.e., $\boldsymbol{\Upsilon} _{0}$. We then present the specific DOA estimation procedure based on this configuration.

The proposed protocol divides the total observation time $T$ into $T_{\textrm{y}}$ blocks, each of length $T_{\textrm{x}}$, so that $T=T_{\textrm{x}}T_{\textrm{y}}$. The phase shift vectors for the first to the $L$-th layers are determined by employing the optimization process described in Section \ref{sec3_2} and remain the same during $T$ snapshots. By contrast, the phase shift vector for the input layer of the SIM is reconfigured at each time slot in order to generate a set of DFT matrices having orthogonal spatial frequency bins. Specifically, at the snapshot $t$, the phase shift of the $n$-th meta-atom imposed on the input layer is configured as:
\begin{align}
 \xi _{0,n,t} =-2\pi \frac{\left ( n_{\textrm{x}}-1 \right )\left ( t_{\textrm{x}}-1 \right )}{N_{\textrm{x}}T_{\textrm{x}}}-2\pi \frac{\left ( n_{\textrm{y}}-1 \right )\left ( t_{\textrm{y}}-1 \right )}{N_{\textrm{y}}T_{\textrm{y}}}, \label{eq22}
\end{align}
where $n_{\textrm{y}}$ and $n_{\textrm{x}}$ are as defined in \eqref{eq17}, $t_{\textrm{y}}$ and $t_{\textrm{x}}$ represent the block index and the time slot index within that block, respectively, which are defined by $t_{\textrm{y}} \triangleq \left \lceil t/T_{\textrm{x}} \right \rceil$ and $t_{\textrm{x}} \triangleq t-\left ( t_{\textrm{y}}-1 \right )T_{\textrm{x}}$, respectively.

Note that upon right-multiplying $\boldsymbol{G}$ (i.e., the well-fitted version of $\boldsymbol{F}$) by $\boldsymbol{\Upsilon }_{0,t}$, the SIM implicitly characterizes a set of 2D DFT matrices whose frequency bins are mutually orthogonal to each other. Under the noiseless received signal model, the EM waves propagating through the optimized SIM are automatically focused on the specific antenna and snapshot indices, returning the on-grid DOA estimate and the spatial frequency offset component of the incoming signal, respectively. As a result, the DOA parameters of the incoming signal can be readily estimated by measuring the energy distribution across the receiving antenna array, which is in contrast to conventional DOA estimation algorithms relying on phase-sensitive receivers and array signal processing.

Specifically, let $r_{n,t}$ represent the signal received at the $n$-th probe in the $t$-th snapshot. After collecting the received signals over $T$ snapshots, we then search for the index of the strongest signal magnitude. The 2D index of the peak is inferred as follows:
\begin{align}
\left [ \hat{n},\hat{t} \right ] = \arg\underset{n=1,2,\cdots ,N,\atop t=1,2,\cdots , T}{\max}\left | r_{n,t} \right |^{2}. \label{eq28}
\end{align}

Therefore, the corresponding electrical angles of the incident signal are obtained by
\begin{align}
 \hat{\psi}_{\textrm{x}} &= \textrm{mod}\left [ 2\left ( \frac{\hat{n}_{\textrm{x}}-1}{N_{\textrm{x}}} +\frac{\hat{t}_{\textrm{x}}-1}{N_{\textrm{x}}T_{\textrm{x}}} \right )+1,2 \right ]-1, \label{eq29}\\
 \hat{\psi}_{\textrm{y}} &=\textrm{mod}\left [ 2\left ( \frac{\hat{n}_{\textrm{y}}-1}{N_{\textrm{y}}} +\frac{\hat{t}_{\textrm{y}}-1}{N_{\textrm{y}}T_{\textrm{y}}} \right )+1,2 \right ]-1, \label{eq30}
\end{align}
respectively, where $\hat{n}_{\textrm{y}}$ and $\hat{n}_{\textrm{x}}$ are defined as in \eqref{eq17}, while $\hat{t}_{\textrm{y}}$ and $\hat{t}_{\textrm{x}}$ are as in \eqref{eq22}.

Based on the estimates in \eqref{eq29} and \eqref{eq30}, the estimated azimuth and elevation angles $\hat{\varphi }$ and $\hat{\vartheta }$ are given by
\begin{align}
 \hat{\varphi }&=\arctan\left ( \frac{\hat{\psi} _{\textrm{y}}d_{\textrm{x}}}{\hat{\psi} _{\textrm{x}}d_{\textrm{y}}} \right ), \label{eq31}\\
 \hat{\vartheta }&=\arcsin\left (\frac{1}{\kappa } \sqrt{\frac{\hat{\psi} _{\textrm{x}}^{2}}{d_{\textrm{x}}^{2}}+\frac{\hat{\psi} _{\textrm{y}}^{2}}{d_{\textrm{y}}^{2}}} \right ). \label{eq32}
\end{align}

In contrast to conventional radar systems, the SIM estimates the DOA by directly processing the received RF signals, without the need for an individual RF chain and an ADC at each antenna element. This substantially reduces both the hardware cost and the energy consumption, which has great potential for onboard UAV applications employing a SIM to probe the DOA of ground targets.

\begin{figure}[!t]
\centering
\includegraphics[width=5.9cm]{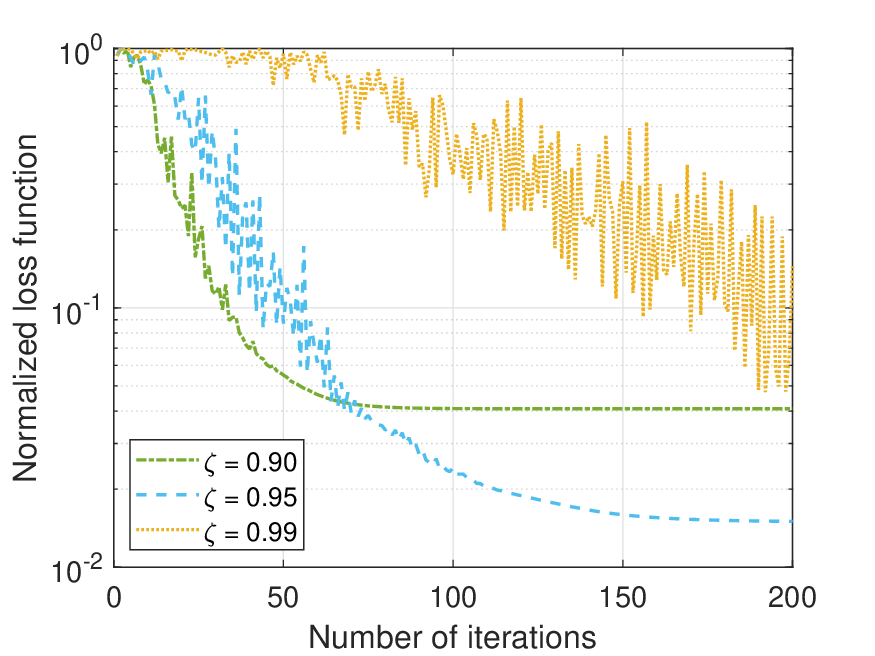}
\caption{The convergence behavior of the proposed gradient descent algorithm.}\vspace{-0.6cm}
\label{fig_2}
\end{figure}

\section{Simulation Results}\label{sec6}
In this section, we conduct numerical simulations to evaluate the performance of the SIM-based DOA estimator. Specifically, we consider the scenario of estimating the DOA of a single source, as shown in Fig. \ref{fig_1}. The system operates at $60$ GHz. In our simulations, we consider $N_{\textrm{x}}=N_{\textrm{y}}=4$. The spacing between the adjacent meta-atoms on the same metasurface in the SIM is set to $d_{\textrm{x}} = d_{\textrm{y}} = \lambda/2$. The receiver antenna array is arranged in the same way as the input layer of the SIM, both with half-wavelength element spacing.

In Fig. \ref{fig_2}, we evaluate the convergence behavior of the proposed gradient descent algorithm for optimizing a nine-layer SIM having the layer spacing of $d_{\textrm{layer}} = \lambda$. Each square metasurface contains $M= 12\times12 = 144$ meta-atoms. Three different decay parameters of $\zeta = 0.9,0.95,0.99$ are considered. It is demonstrated in Fig. \ref{fig_2} that as the iterations proceed, the proposed gradient descent method would gradually converge for a moderate value of the decay parameter, such as $0.95$. For a value of $\zeta$ close to $1$, the algorithm may overshoot frequently and require more iterations to converge. Additionally, while the gradient descent may reduce the loss function rapidly at the early stage of less than $50$ iterations for $\zeta = 0.9$, it may get trapped in a locally optimal point. As a result, the decay parameter of $0.95$ performs the best after running $200$ iterations.

\begin{figure}[!t]
\centering
\includegraphics[width=5.9cm]{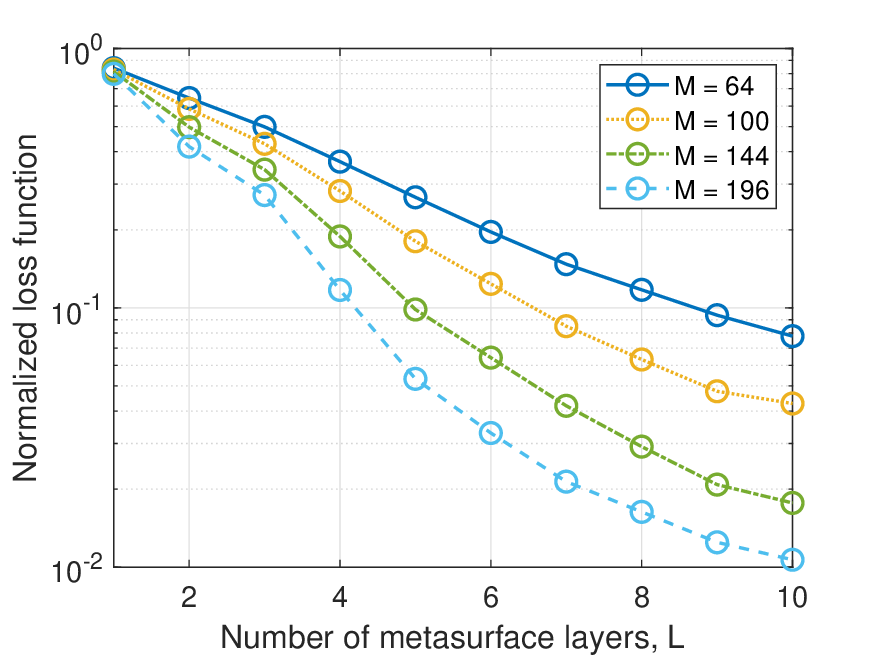}
\caption{The normalized loss function versus the number of metasurface layers.}\vspace{-0.6cm}
\label{fig_3}
\end{figure}

Furthermore, we evaluate the ability of the SIM to mimic the 2D DFT versus the number of metasurface layers $L$. Specifically, we consider four setups, namely $M = 64, 100, 144, 196$ meta-atoms on each layer. For each setup, the SIM phase shifts are optimized using the gradient descent method with a decay parameter of $0.95$ over a maximum of $200$ iterations. The simulation results of Fig. \ref{fig_3} demonstrate that a SIM having few layers is unable to fit the 2D DFT matrix well. As the number of layers $L$ increases, the SIM succeeds in better approximating the 2D DFT in the wave-domain. Further increasing the number of layers may cause the loss function to level off. Additionally, it is demonstrated that the fitting performance also improves, as the number of meta-atoms $M$ on each layer increases. Specifically, observe from Fig. \ref{fig_3} that increasing $M$ from $64$ to $196$ improves the normalized fitting MSE by nearly an order of magnitude.

\begin{figure}[!t]
\centering
\includegraphics[width=5.9cm]{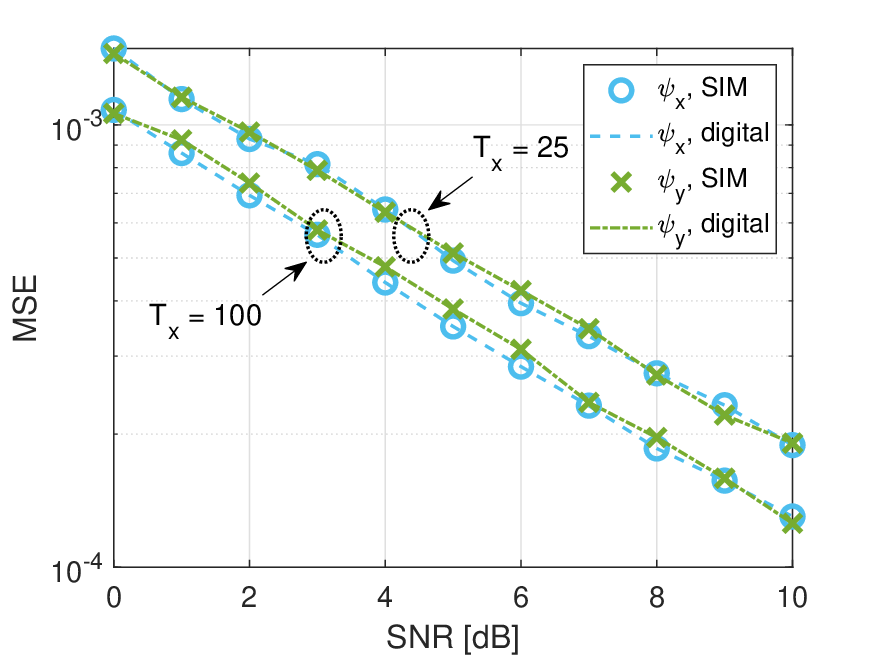}
\caption{MSE versus SNR.}\vspace{-0.6cm}
\label{fig_4}
\end{figure}

Next, we examine the performance of the proposed SIM-based DOA estimator by evaluating its MSE versus the SNR, while the SIM parameters are kept the same as in Fig. \ref{fig_2}. For brevity, we consider the electrical angels $\psi _{\textrm{x}}$ and $\psi _{\textrm{y}}$ instead of the physical angles, which are both uniformly distributed in $\left [ -1,1 \right ]$. The simulation results shown in Fig. \ref{fig_4} are obtained by averaging $100$ independent experiments. More specifically, we consider two different numbers of snapshots, namely $T_{\textrm{x}} = T_{\textrm{y}} = 25$ and $100$. For comparison, we also plot the estimation MSE using the proposed estimation protocol associated with the digital 2D DFT. As expected, the MSE is improved as the SNR increases. Specifically, the MSE improves by $10$ dB for every $10$ dB increase in SNR. Additionally, increasing the number of snapshots per block from $T_{\textrm{x}} = 25$ to $T_{\textrm{x}} = 100$ provides an extra $2$ dB performance gain, thanks to the finer granularity. It is demonstrated that at the SNR of $10$ dB, the proposed SIM-based channel estimator achieves an MSE of approximately $10^{-4}$ when using $T_{\textrm{x}} = 100$ snapshots in each block. Moreover, under all setups, our SIM achieves comparable performance to that of its digital counterpart.

\begin{figure}[t!]
	\centering
	\subfloat[\label{fig11a}Case I: $\bar \psi_{\textrm{x}}=-0.67$, $\bar \psi_{\textrm{y}}=-0.48$;]{
		\includegraphics[width=4cm]{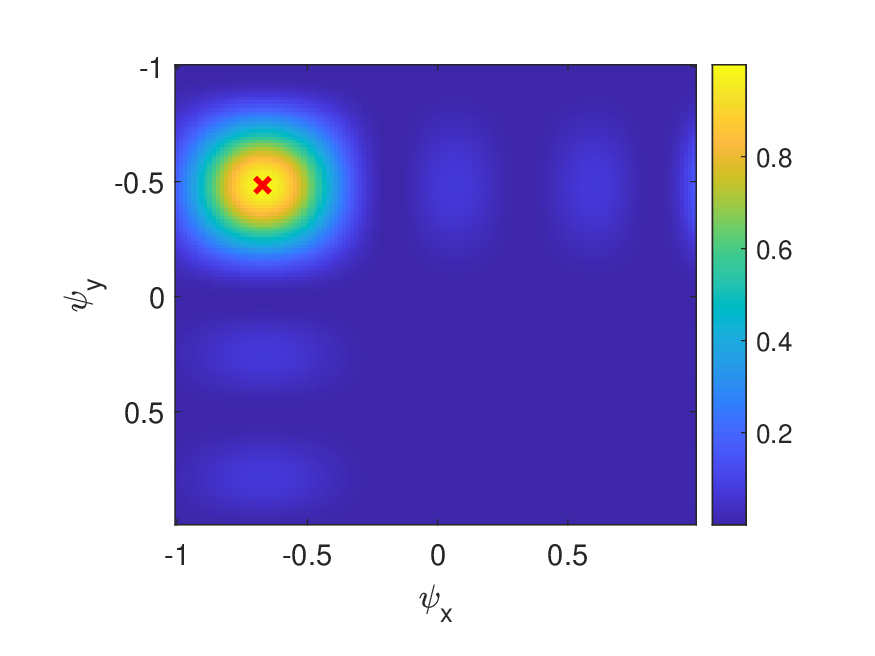}}
	\subfloat[\label{fig11b}Case II: $\bar \psi_{\textrm{x}}=0.53$, $\bar \psi_{\textrm{y}}=-0.34$;]{
		\includegraphics[width=4cm]{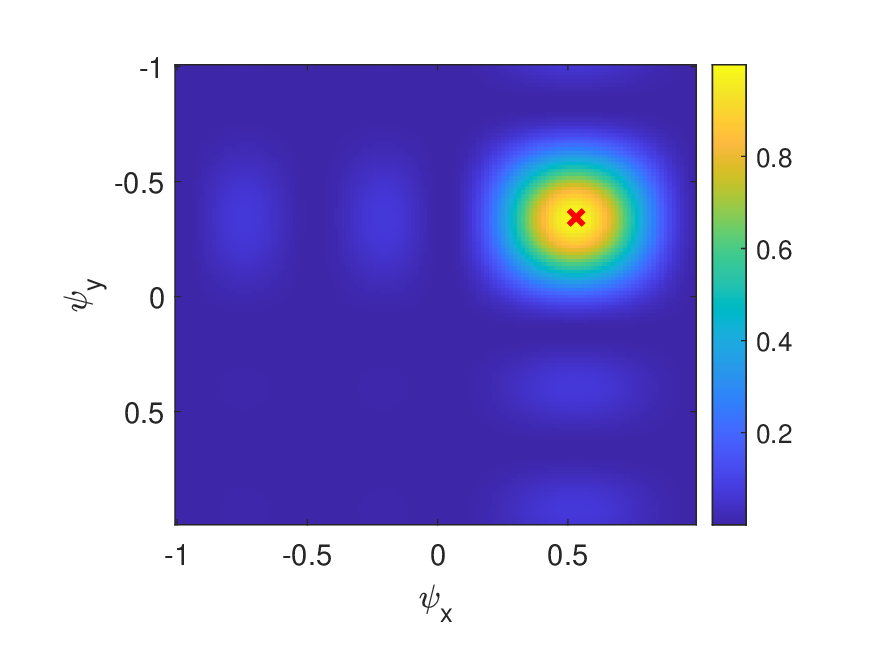}}
	\\
	\subfloat[\label{fig11c}Case III: $\bar \psi_{\textrm{x}}=-0.52$, $\bar \psi_{\textrm{y}}=0.41$;]{
		\includegraphics[width=4cm]{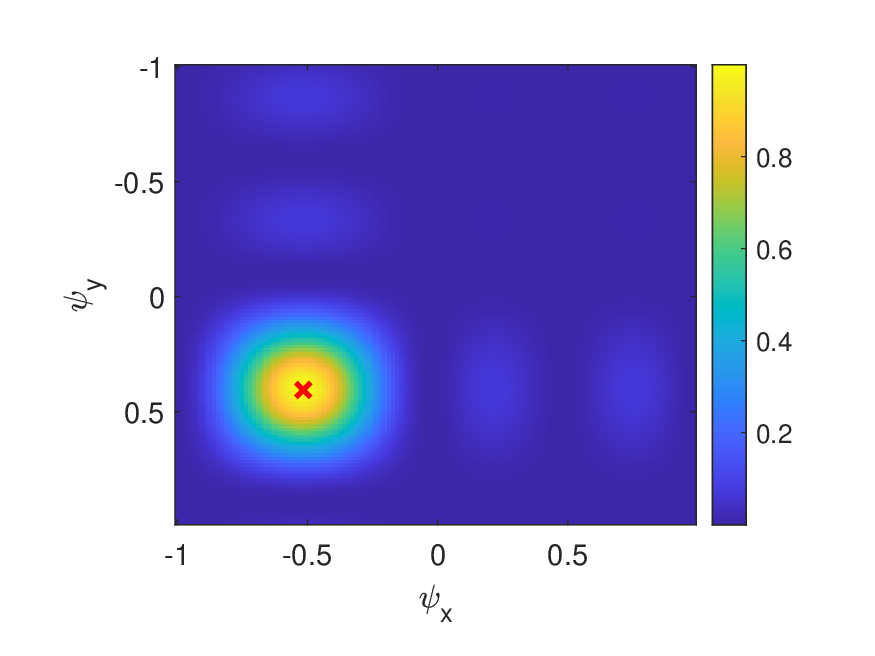}}
	\subfloat[\label{fig11d}Case IV: $\bar \psi_{\textrm{x}}=0.44$, $\bar \psi_{\textrm{y}}=0.33$;]{
		\includegraphics[width=4cm]{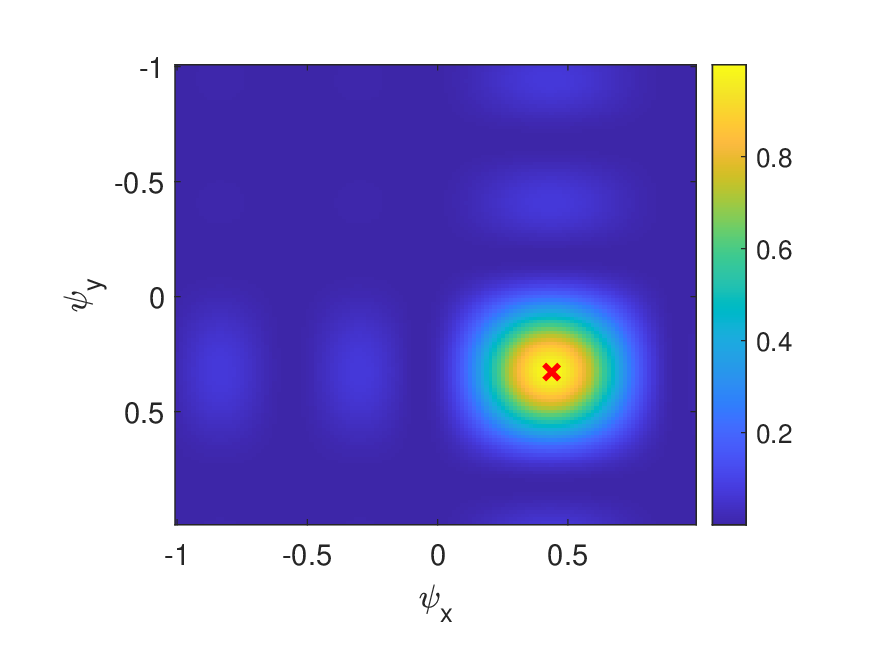}}
	\caption{The spatial spectrum of the incident signal after passing through the SIM.}\vspace{-0.6cm}
	\label{fig_5} 
\end{figure}

Finally, we verify the 2D DFT capability of the SIM by analyzing the spatial-domain spectrum under four different setups. The spectrum is obtained by laminating the outputs of $T_{\textrm{x}} \times T_{\textrm{y}} = 1024$ 2D DFTs over $\left ( 4,4 \right )$ grid points. We first consider a random setup of $\bar \psi_{\textrm{x}}=-0.67$ and $\bar \psi_{\textrm{y}}=-0.48$, marked by the red cross. Observe from Fig. \ref{fig_5}(a) that the highest energy matches both the corresponding antenna as well as the snapshot index. This means that the SIM could return a perfect estimate of the DOA parameters under noiseless conditions. Similarly, the spatial spectra of the other three cases are shown in Figs. \ref{fig_5}(c)-(d), where the energy peak is at the position corresponding to the incident signal's DOA. In a nutshell, the SIM results in a fundamental DOA estimation paradigm shift by directly observing the spatial-domain spectrum instead of the array signal.

\section{Conclusions}\label{sec7}
We proposed a novel SIM architecture for estimating the 2D DOA parameters. By appropriately training the SIM, the spatial EM waves can be directly transformed into their spatial frequency domain as they propagate through the SIM. Furthermore, we designed a protocol to generate a spatial spectrum having orthogonal spatial frequency bins. Thus, one can easily read the DOA by searching for the index having the highest magnitude. Our simulation results indicate that the proposed SIM-based DOA estimator achieves an MSE of $10^{-4}$ under moderate conditions, while allowing for a substantial enhancement in the computation speed at a moderate hardware complexity. As the first attempt in this area, we considered a single source in this paper, while DOA estimates of multiple sources using the SIM require future research efforts \cite{arXiv_2023_An_Toward}.

\bibliography{ref}
\bibliographystyle{IEEEtran}
\end{document}